\documentclass[10pt,journal,compsoc]{IEEEtran}
\ifCLASSOPTIONcompsoc
  \usepackage[nocompress]{cite}
\else
  \usepackage{cite}
\fi

\usepackage{graphicx}

\ifCLASSINFOpdf
\else
\fi

\usepackage{multirow}
\usepackage{hhline}
\usepackage{array}
\usepackage{amsmath}
\usepackage{amssymb}
\usepackage{bm}
\usepackage{color}
\usepackage[colorlinks = true,
            linkcolor = blue,
            urlcolor  = blue,
            anchorcolor = blue]{hyperref}
\definecolor{COLOR}{rgb}{1.0, 0.25, 0.25}

\begin{document}
%
\title{A Multi-Stage Attentive Transfer Learning Framework for Improving COVID-19 Diagnosis}
%
%
%
%

\author{Yi~Liu,
        and~Shuiwang~Ji,~\IEEEmembership{Senior Member,~IEEE}
\thanks{Yi Liu and Shuiwang Ji are with Department
of Computer Science \& Engineering, Texas A\&M University, College Station, TX 77843, USA
(email: yiliu@tamu.edu, sji@tamu.edu).}}

\IEEEtitleabstractindextext{%
\begin{abstract}
Computed tomography (CT) imaging is a promising approach to
diagnosing the COVID-19. Machine learning methods can be employed to
train models from labeled CT images and predict whether a case is
positive or negative. However, there exists no publicly-available
and large-scale CT data to train accurate models. In this work, we
propose a multi-stage attentive transfer learning framework for
improving COVID-19 diagnosis. Our proposed framework consists of
three stages to train accurate diagnosis models through learning
knowledge from multiple source tasks and data of different domains.
Importantly, we propose a novel self-supervised learning method to
learn multi-scale representations for lung CT images. Our method
captures semantic information from the whole lung and highlights the
functionality of each lung region for better representation
learning. The method is then integrated to the last stage of the
proposed transfer learning framework to reuse the complex patterns
learned from the same CT images. We use a base model integrating
self-attention (ATTNs) and convolutional operations. Experimental
results show that networks with ATTNs induce greater performance
improvement through transfer learning than networks without ATTNs.
This indicates attention exhibits higher transferability than
convolution. Our results also show that the proposed
self-supervised learning method outperforms several baselines
methods.

\end{abstract}

\begin{IEEEkeywords}
COVID-19 diagnosis, transfer learning, self-supervised learning,
attention, transferability, medical image computing.
\end{IEEEkeywords}}

\maketitle

\IEEEdisplaynontitleabstractindextext

%
\IEEEpeerreviewmaketitle

\IEEEraisesectionheading{\section{Introduction}\label{sec:introduction}}

%
%
%
%
\IEEEPARstart{T}{he} COVID-19 pandemic has spread rapidly and infected millions of people
worldwide. A critical step to fight against the spreading of COVID-19 is effective
diagnosis of the infected cases~\cite{tang2020laboratory,udugama2020diagnosing}.
The commonly used approach for COVID-19 diagnosis is reverse
transcription polymerase chain reaction (RT-PCR).
It usually takes hours for results and test kits are in great shortage in some counties and areas~\cite{xie2020chest,zhou2020clinical}.
Computed tomography (CT)-aided diagnosis has become a weighty alternative due to its nature of wide availability and easy accessibility~\cite{bernheim2020chest,xie2020chest,li2020coronavirus}. To accelerate the reading of CT images, machine learning approaches have been employed
to learn patterns from labeled images and then automatically make prediction for any newly
obtained CT image~\cite{wang2020covid,gozes2020rapid,alimadadi2020artificial,wynants2020prediction}.

A major challenge of CT-aided automatic diagnosis is the lack of labeled data~\cite{wang2020covid,wang2020deep,he2020sample}.
To date, the largest labeled CT dataset~\cite{he2020sample} that is publicly available only contains several hundred images.
Models trained on such small-scale datasets may generate unsatisfactory prediction results for newly obtained CT images.
It is natural to leverage transfer learning to train more powerful models for accurate COVID-19 diagnosis.
Transfer learning is a popular machine learning technique that learns a model from a source task
where the labeled data is sufficient and transfers the learned knowledge
to a target task~\cite{pan2009survey,weiss2016survey,yosinski2014transferable}.
Existing work~\cite{he2020sample,wang2020covid} mainly focuses on using pretrained deep neural networks (DNNs) to improve the prediction performance
for the target task of COVID-19 prediction. However, there is no work examining which network component, such as a convolutional layer or an attention layer, can induce
larger performance improvement through transfer learning.
In addition, there is little work presenting a unified transfer learning framework for
medical image analysis, especially for COVID-19 CT image prediction.

In this work, we propose a multi-stage attentive transfer learning framework for improving CT-based COVID-19 diagnosis.
Our proposed framework is composed of three stages based on the machine learning approaches and data used in source tasks.
First, we perform supervised transfer learning from natural images (STL-N) and supervised transfer learning from medical images (STL-M)
to learn knowledge from large-scale labeled natural images and medical images, respectively.
After that, we design a novel self-supervised task and perform self-supervised transfer learning from medical images (SSTL-M)
to extract complex patterns from the used medical CT images.
For the networks used in the transfer learning framework,
we integrate self-attention layers (ATTNs) into convolutional neural networks (CNNs) such as ResNets
to compare transferability of convlotional layers and ATTNs. Specifically, We categorize networks into two groups,
one of which contains ResNets and the other contains the same ResNets with ATTNs inserted in.
Then both groups are pretrained on the same source tasks and data to compare transferability of the two groups.

We perform self-supervised transfer learning in the last stage of the proposed transfer learning framework.
Existing self-supervised methods~\cite{doersch2015unsupervised,gidaris2018unsupervised,lee2019rethinking,kolesnikov2019revisiting,zhai2019s4l}
achieve superior results on natural images but usually generate poor predictions on medical images.
By referring to biological domain knowledge of substructures of the human lung,
we design a novel self-supervised method to learn multi-scale representations for lung CT images.
Our method is capable of learning
representations at both the image level and the region level.
By doing this, sufficient semantic information from the whole lung is captured and the functionality of
each lung region is highlighted for better representation learning.
Then self-supervised transfer learning is performed
to reuse the complex inherent patterns learned
from the same CT images to improve the performance of the target task.

We conduct extensive experiments to evaluate our proposed approach.
Experimental results show that after pretrained with our proposed multi-stage transfer learning framework,
networks with ATTNs
achieve much better performance for CT-aided COVID-19 prediction compared with the baseline ResNets.
This indicates the effectiveness of integrating ATTNs into our transfer learning framework.
More importantly, it is shown that networks with ATTNs
result in much
larger performance improvement through transfer learning compared with
convolutional layers. This points out that compared with convolution, attention can transfer knowledge from source tasks to target tasks more easily,
which essentially
reveals that attention exhibits higher transferability than convolution.
In addition, we show that our proposed self-supervised learning method achieves best performance
compared with several SOTA baselines.
This demonstrates the effectiveness of our method of
learning multi-scale representations
of lung CT images and highlighting the functionality of
each lung region.
Our qualitative results demonstrate that using attention for transfer learning
can successfully detect important regions for prediction.
Overall, our major contributions are
summarized as follows:
\begin{itemize}
    \item We propose a multi-stage attentive transfer learning framework for improving CT-aided COVID-19 diagnosis.
    The proposed framework successfully learns and transfers knowledge from multiple source tasks and data of different domains
    for accurate COVID-19 diagnosis.
    \item We propose a novel self-supervised learning method for medical images.
    Our method enables multi-scale representation learning for lung CT images and outperforms existing self-supervised learning methods.
    \item We not only show that networks with attention layers are more powerful through transfer learning,
    but also demonstrate that attention has higher transferability than convolution.
    To our best knowledge, this is the first work to compare transferability of attention and convolution.

\end{itemize}

\section{Related Work}
In this section, we introduce related work of transfer learning and attention mechanism.
\subsection{Transfer Learning}
Transfer learning aims at transferring knowledge across different tasks~\cite{pan2009survey,yosinski2014transferable}.
Generally, it learns knowledge from a source task and transfers
the knowledge to a target task. In practice, it is usually difficult
to collect sufficient training data for a target task. Training a model
on insufficient data may result in unsatisfactory prediction results. Transfer
learning is used to first train a model on a source task where the training
data is sufficient. Then the pretrained model serves as the starting point and is
finetuned on the target task. For instance, in visual recognition, a model
is usually trained on ImageNet that contains millions of labeled training samples.
After that, the trained weights are used as initial weights for downstream tasks
such as semantic segmentation and objective detection. Transfer learning has achieved
success across various artificial intelligence domains, including natural language processing~\cite{devlin2018bert}
computer vision~\cite{ren2015faster}, and biomedical image analysis~\cite{raghu2019transfusion,esteva2017dermatologist}.

There exist several categorization criteria of transfer learning~\cite{pan2009survey,weiss2016survey}.
Based on the machine learning approaches used in the source task,
transfer learning could be categorized into supervised transfer learning, unsupervised transfer learning,
and semi-supervised transfer learning, etc. Generally, supervised and semi-supervised learning have been studied intensively,
and unsupervised learning is a promising research area as labeling is usually expensive.
Self-supervised learning is a type of unsupervised learning strategy that has gained more and more popularity
recently~\cite{dosovitskiy2015discriminative,doersch2015unsupervised,gidaris2018unsupervised}.
It aims at supervised feature learning where the supervision is provided by the data.
The supervised tasks are the key for self-supervised learning.
Earlier work for supervised tasks on images basically predicts positions or context
for a local patch~\cite{doersch2015unsupervised,gidaris2018unsupervised}.
Recent work~\cite{he2020momentum,chen2020improved,chen2020simple} mainly performs two random sets of data augmentations on
a pair of images and predicts whether the two images are the same or not.
The contrastive loss is commonly used for these methods.
The input for the contrastive loss contains a query vector $\bm{x}_q$ from an image $\bm{X}$, a key vector $\bm{x}_{k_+}$ from the same image $\bm{X}$, and another $n$
key vectors from $n$ images that are different from $\bm{X}$.
Then the contrastive loss is essentially a log-loss of a $(n + 1)$-way softmax classifier that tries to
classify $\bm{x}_q$ to $\bm{x}_{k_+}$ rather than the other $n$
key vectors.

\begin{figure*}[!th]
\centering
\includegraphics[width=0.8\textwidth]{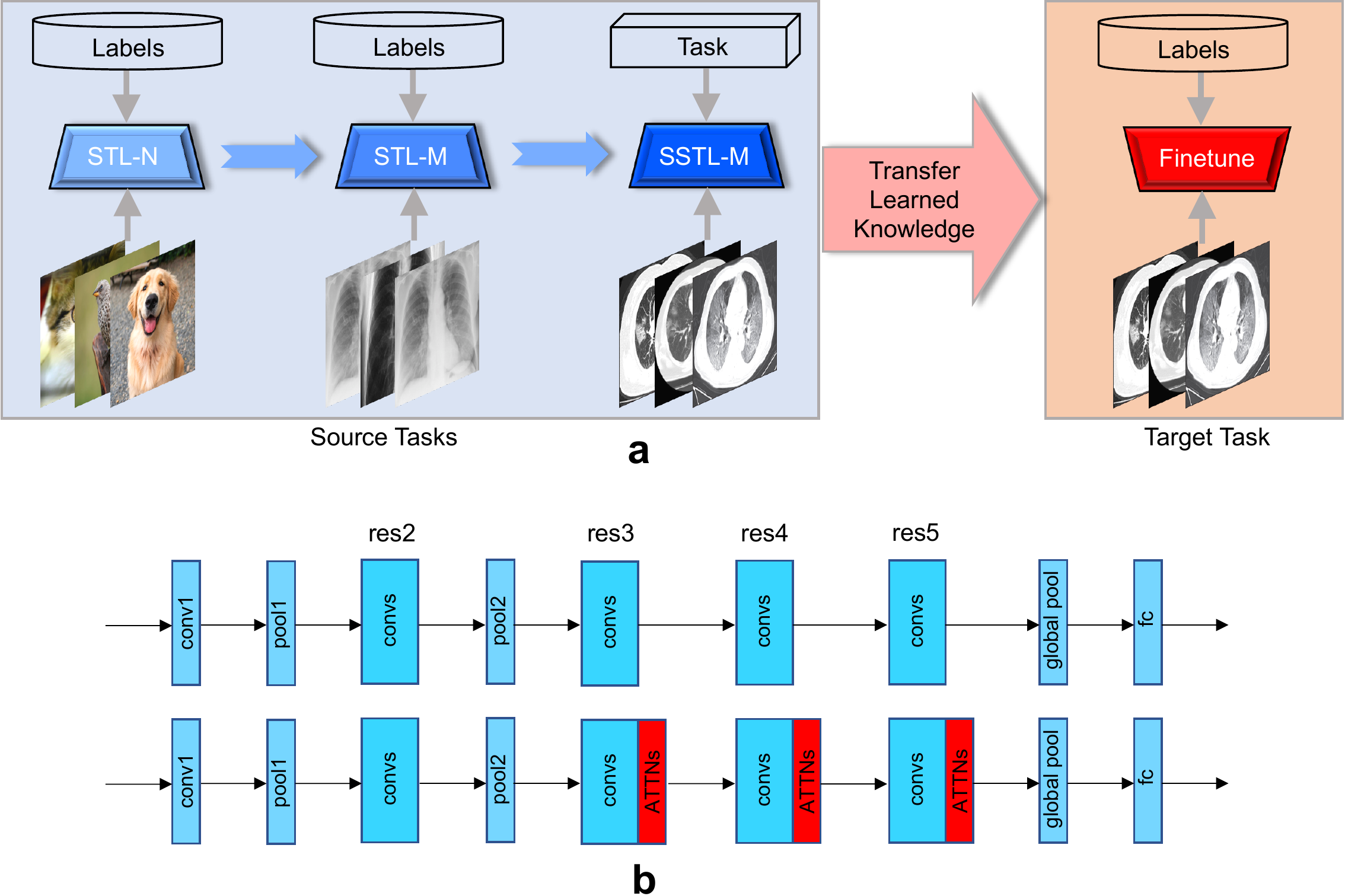}
\caption{\textbf{The proposed multi-stage transfer learning framework and network settings.}
\textbf{a}. Our proposed multi-stage transfer learning framework that contains three
source tasks, these are,
supervised transfer learning from natural images (STL-N), supervised transfer learning from medical images (STL-M),
and self-supervised transfer learning from medical images(SSTL-M).
Then the learned knowledge is transferred to the target task of COVID-19
diagnosis based on CT images.
There exist a large amount of labeled data in the STL-N and STL-M.
In the SSTL-M, we carefully design a task based on the data to
extract complex inherent patterns from medical images.
\textbf{b}. Two groups of network settings. The top one represents a standard ResNet
architecture. The bottom one illustrates a ResNet with self-attention layers (ATTNs)
inserted in the $\mbox{res}_3$, $\mbox{res}_4$ and $\mbox{res}_5$ blocks.
} \label{fig:framework}
\end{figure*}

\subsection{Attention Mechanism}
In this section, we describe the attention mechanism,
which
captures long-range dependencies from input~\cite{wang2018non,vaswani2017attention}.
Given the input tensor $\mathcal{X}\in\mathbb{R}^{h\times w\times c}$,
the attention mechanism first performs $1\times 1$ convolution three times
and achieves three tensors; those are, the query $\mathcal{Q}\in\mathbb{R}^{h\times w\times d_k}$,
the key $\mathcal{K}\in\mathbb{R}^{h\times w\times d_k}$ and the value $\mathcal{V}\in\mathbb{R}^{h\times w\times d_v}$.
These tensors are unfolded into three matrices along mode-3~\cite{kolda2009tensor}, resulting in
$Q\in\mathbb{R}^{d_k\times hw}$, $K\in\mathbb{R}^{d_k\times hw}$ and
$V\in\mathbb{R}^{d_v\times hw}$, respectively.
Then the intermediate output is computed as
\begin{equation}\label{eq:total_o}
O = V\times\mbox{Softmax}(K^TQ)\in\mathbb{R}^{d_v\times hw},
\end{equation}
where $\mbox{Softmax}(\cdot)$ is performed on columns
such that every column sums to 1.
Finally, the obtained matrix $O$ is converted back
to a tensor $\mathcal{O}\in\mathbb{R}^{h\times w\times d_v}$,
as the final output of the attention mechanism.

Essentially, $(K^TQ)$ generates a matrix of sizes $hw \times hw$,
which can be treated as $hw$ attention heatmaps. Each heatmap
contains $hw$ attention weights, and $O$ is computed as a weighted sum of all vectors in $V$.
To this end, the response at each position of the output $\mathcal{O}$ is
dependent on all positions of $\mathcal{V}$, which is
just achieved by performing a linear transformation on the input $\mathcal{X}$.
As a result. long-range dependencies from the input are captured by attention.
Notably, the output of attention is input dependent. Different from convolution
where weights are learnable parameters, attention weights are computed from the input.

\section{Multi-Stage Attentive Transfer Learning Framework}
In this section, we introduce our proposed multi-stage attentive transfer learning framework
and network settings.

\subsection{Framework Overview} \label{sec:stl}
Given a new CT image, our objective is to predict whether it's COVID-19 positive or negative
 based on the trained model.
However, existing COVID-19 CT data is small-scale and insufficient to train a powerful model,
which usually leads to poor prediction performance.
It's natural to leverage transfer learning to achieve more powerful models
and boost the performance of COVID-19 prediction.
An illustration of our proposed multi-stage transfer learning framework is provided in
Figure~\ref{fig:framework}a. Specifically,
we first conduct two supervised source tasks,
namely supervised transfer learning from natural images (STL-N) and supervised transfer learning from medical images (STL-M),
to learn models from large-scale labeled data for the target task. After that, we perform
self-supervised transfer learning from medical images (SSTL-M) to learn complex inherent patterns from the used CT images.

\subsection{Network Settings} \label{sec:network}

For the networks used in our transfer learning framework, we categorize them into two groups to compare transferability
between self-attention layers (ATTNs) and convolutional layers.
The first group contains standard CNNs such as ResNet-50 and ResNet-101.
For the other group,
we follow the settings in the work~\cite{wang2018non} and insert ATTNs
in these backbone ResNets.
An illustration of the network settings is provided in Figure~\ref{fig:framework}b.
Generally, there exist four residual blocks in
the family of ResNets, namely $\mbox{res}_2$, $\mbox{res}_3$, $\mbox{res}_4$ and $\mbox{res}_5$, respectively~\cite{he2016deep}.
It is shown that networks with ATTNs inserted in the $\mbox{res}_3$ and $\mbox{res}_4$ obtain the best performance~\cite{wang2018non}.
We use similar strategies and insert most ATTNs in the $\mbox{res}_3$ and $\mbox{res}_4$. In addition, we propose to insert another
ATTN in the $\mbox{res}_5$ before the global average pooling to qualitatively demonstrate the effectiveness of ATTNs in transfer learning.
Then both the groups of networks are applied to our
multi-stage transfer learning framework
to compare transferability of the two groups.
Essentially, networks with and without ATTNs
are pretrained on the same source tasks to compare
which of ATTNs and
convolutional layers can transfer knowledge from these source tasks more easily.

\section{Supervised Transfer Learning}
The labeled CT data for COVID-19 diagnosis is limited.
Training networks directly on these CT images may result in poor
performance for COVID-19 detection.
There exist large-scale labeled datasets from other domains or diseases.
We use supervised transfer learning to learn and transfer knowledge from
these labeled data to facilitate the CT-aided COVID-19 diagnosis.

We first perform STL-N on ImageNet,
a large-scale collection of natural images and the most popular labeled dataset for model pretraining.
Both the groups of networks introduced in Section~\ref{sec:network}
are pretrained on ImageNet to learn knowledge from
natural images.
When applied to new tasks such as COVID-19 CT image prediction, the transferability of two
different categories can be estimated by the performance improvement induced by transfer learning.
Notably, networks with and without ATTNs
are pretrained on ImageNet to compare the transferability
of ATTNs and
convolution layers from natural images.

It is obvious that nature images and medical images (such as CT images) follow different distributions. Pretraining on natural images enables
models to learn common patterns shared by natural and medical images,
but fails to learn distinguishing patterns for medical images.
Hence, we conduct STL-M to pretrain models on
existing large-scale labeled medical images. Even though labeled CT images for COVID-19 diagnosis are scarce,
there exist redundant sources for annotated medical images from other domains,
such as chest X-ray (CXR) images for COVID-19, or CT images for regular pneumonia.
Pretraining on these medical images enables models to learn inherent patterns in medical images and extract strong
features for accurate COVID-19 diagnosis. Similar to STL-N, the two groups of networks
that with and without ATTNs are both pretrained on the labled medical images to compare transferability
of these two groups.
Notably, by performing the two stages of supervised transfer learning STL-N and STL-M, we
compare transferability of ATTNs and convolutional layers on two scenarios, where
the source data follows
different or similar distributions with the target data.

\section{Self-supervised Transfer Learning} \label{sec:sstl}
We proposed to transfer knowledge from labeled natural and medical images by
performing supervised transfer learning STL-N and STL-M. However, there still exists divergence in
distributions between the labeled source data and the target data. Notably, medical
images from other modalities or diseases still have a domain shift from the lung CT images for COVID-19.
Hence, we perform self-supervised transfer learning as the last stage in our framework
to obtain knowledge from
the same CT images.
Existing self-supervised learning methods achieve good performance on natural images but usually
result in poor performance on medical images (like CT images).
In this section, we introduce a novel self-supervised learning method for medical images and transfer
the learned knowledge to COVID-19 detection.
We denote this stage as self-supervised transfer learning from medical images (SSTL-M),
as illustrated in Figure~\ref{fig:framework}a.
The objective of SSTL-M is to learn complex and inherent patterns from CT images
by performing a carefully-designed self-supervised task.

\subsection{Method Overview}
It is vital to design an appropriate source
task to obtain redundant information from the CT images. Currently, self-supervised learning methods,
including MoCo (v1~\cite{he2020momentum} and v2~\cite{chen2020improved})
and simCLR~\cite{chen2020simple}, have achieved the SOTA performance on tasks for natural images,
but usually result in poor performance on tasks related to medical images~\cite{chen2019self}.
Essentially, these methods apply two sets of random data augmentations on the same image
then force the network make a positive prediction.  A positive pair contains two same images (one
image actually). Negative pairs are added where a negative pair contains two
different images. By doing this, inherent patterns from input images are learned
and can be transferred to other target tasks. However, the data augmentation methods
used in these tasks are commonly used techniques (such as rotation, flipping) and may not be
strong enough to extract semantic patterns from medical images. In addition,
medical images such as lung CT images are usually symmetric in structure.
Applying data augmentation on the whole image may fail to extract distinguishing features
from a specific region.

\begin{figure*}[!th]
\centering
\includegraphics[width=\textwidth]{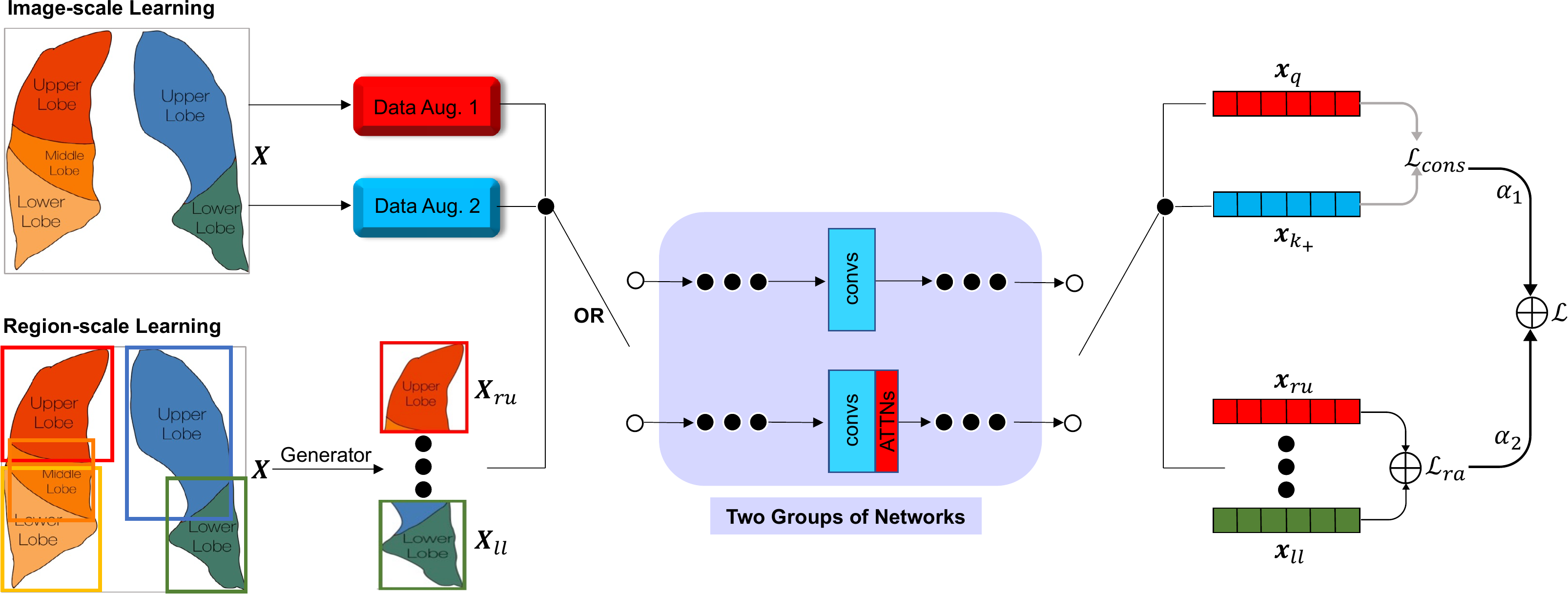}
\caption{\textbf{The proposed self-supervised transfer learning method.}
The method is composed of the image-scale learning branch and the region-scale learning branch.
For the purpose of simplicity, we only illustrate the scenario
for a positive pair based on an input lung CT image $\bm{X}$ but omit
the scenario for negative pairs.
For image-scale learning, two sets of random data augmentations
are applied on $\bm{X}$, denoted as data aug. 1 and data aug. 2. This results in two augmented images.
For the region-scale learning, five regions, including $\bm{X}_{ru}$, $\bm{X}_{rm}$, $\bm{X}_{rl}$, $\bm{X}_{lu}$ and $\bm{X}_{ll}$
and generated from the image $\bm{X}$ by the Generator operator as defined in Equation~\ref{eq:rsl_input}.
Generator is composed of three operators, these are, the locate operator, the crop operator and the resize operator.
All of the intermediate outputs of the two branches are passed to the same network.
We have two groups of networks as introduced in Section~\ref{sec:network} and Figure~\ref{fig:framework}b. The switch
OR means that we choose either of the two groups for one experiment.
And we conduct experiments on both the group to compare the transferability of
ATTNs and convolutional layers.
In image-scale learning, the networks input two augmented images and output two reprsenantions
$\bm{x}_q$ and $\bm{x}_{k_+}$, based on which the contrastive loss $\mathcal{L}_{cons}$ (defined in Equation~\ref{eq:contra1}) could be computed.
Region-scale learning is a multi-task learning task to predict the predefined
class of each region.
The region-aware loss $\mathcal{L}_{ra}$ (defined in Equation~\ref{eq:ra_loss}) for the image $\bm{X}$ is simply the sum of the cross-entropy losses of the five regions.
The final loss $\mathcal{L}$ is the weighted
sum of the contrastive loss $\mathcal{L}_{cons}$ of the image-scale learning and $\mathcal{L}_{ra}$ of
the region-scale learning. Note again that we neglect negative pairs for clear illustration, and
the version for $\mathcal{L}$ considering negative pairs
is defined in Equation~\ref{eq:final}.
} \label{fig:ssl}
\end{figure*}

In this work,
we propose a novel self-supervised learning method
to learn multi-scale representations for lung CT images, as illustrated in Figure~\ref{fig:ssl}.
Our proposed method is composed of two branches, these are,
image-scale representation learning for the whole lung structure, and region-scale
learning for different substructures of a lung. We use a contrastive self-supervised pipeline~\cite{he2020momentum} for the former.
For the latter, we design a task referring to prior domain
knowledge based upon
biological structures of the human lung.
Specifically, humans have two lungs, a right lung, and a left lung.
The right lung has three lobes, the upper lobe,
the middle lobe, and the lower lobe.
The left lung is a little smaller and composed of
two lobes, the upper lobe and the lower lobe~\cite{drake2009gray}.
These lobes play different roles in biological processes.
For some diseases, infection of a specific lobe can serve as an
important indicator for medical diagnosis~\cite{drake2009gray,bernheim2020chest}.
Hence, it's important to learn inherent patterns for each specific
lobe and highlight structural divergence among all lobes.

\subsection{Image-scale Learning} \label{sec:imagesl}
We first perform image-scale learning based on the whole CT image.
A contrastive self-supervised learning framework is used,
an input sample of which contains a CT image
 $\bm{X}$ to form a positive pair,
and a set of different CT images $\{\bm{X}^i|i=1,...,n\}$ to form negative pairs.
Specifically, two sets of random augmentations are applied on the same image $\bm{X}$,
which results in two images $\bm{X}_q$ and $\bm{X}_k$. $\bm{X}_q$ is passed to
the query encoder to generate a query vector $\bm{x}_q$,
and $\bm{X}_k$ is passed to
the key encoder to generate a key vector $\bm{x}_{k_+}$.
In addition, images $\{\bm{X}^i|i=1,...,n\}$ are also
taken by the same key encoder to obtain a set of key vectors
$\{\bm{x}^i_{k_-}|i=1,...,n\}$. As a result,
$(\bm{x}_q, \bm{x}_{k_+})$ forms a positive pair and
$(\bm{x}_q, \bm{x}^i_{k_-})$ form $n$ negative pairs for the
contrastive loss, which is expressed as
\begin{equation}\label{eq:contra1}
\mathcal{L}_{cons} = -\mbox{log}\frac{\mbox{exp}(\bm{x}_q \cdot \bm{x}_{k_+}/\tau)}{\mbox{exp}(\bm{x}_q \cdot \bm{x}_{k_+}/\tau)+\sum^n_{i=1}\mbox{exp}(\bm{x}_q \cdot \bm{x}^i_{k_-})},
\end{equation}
where $\tau$ is a temperature parameter. Intuitively,
$\mathcal{L}_{cons}$ is the log-loss of a $(n+1)$-way softmax
classifier whose input is $\bm{x}_q$ and the correct label is $\bm{x}_{k_+}$.

Generally, the query encoder and the key encoder share the same network architecture,
but differ in weight update.
The weights of the query
encode are updated by back-propagation.
Different from the query encoder, the key encoder takes as input a large set of images $\{\bm{X}^i|i=1,...,n\}$,
which usually makes it intractable to update the weights by back-propagation.
The MoCo~\cite{he2020momentum,chen2020improved} adopts a momentum update strategy
for the key encoder as
\begin{equation}\label{eq:momentum}
m\theta_k+(1-m)\theta_q\rightarrow\theta_k,
\end{equation}
where $m$ is the momentum parameter that usually takes a large number such as 0.999,
$\theta_k$ and $\theta_q$ denote the weights of the key encoder and
the query encoder, respectively.
Notably, we also use two groups of networks for the query and key encoders
to compare the transferability of
ATTNs and convolutional layers.

\subsection{Region-scale Learning}
We propose to learn distinguishing patterns in five lobes in region-scale learning.
From this perspective,
we carefully design a task to predict the positions of all the five lobes.
Given an input lung CT image $\bm{X}$, we first generate
five regions that cover the five lobes as
\begin{equation}\label{eq:rsl_input}
\bm{X}_{ru},\bm{X}_{rm},\bm{X}_{rl},\bm{X}_{lu},\bm{X}_{ll}  = \mbox{Generator}(\bm{X}),
\end{equation}
where $\bm{X}_{ru},\bm{X}_{rm},\bm{X}_{rl},\bm{X}_{lu},\bm{X}_{ll}$ denote
the regions cover the right lung upper lobe, the right lung middle lobe,
the right lung lower lobe, the left lung upper lobe, the left lung lower lobe, respectively,
Generator denotes a composition of operators to generate the above five region from an input lung CT image.

Generator is composed of three operators, including the locate, crop and resize operators.
The locate operator contains three steps.
First, we compute a location tuple $(x_i, y_i, w_i, h_i)$ for each region $i\in\{ru, rm, rl, lu, ll\}$ from Figure~\ref{fig:ssl},
where $x_i$ and $y_i$ denote the coordinates of the center point of the region $i$, $w_i$ and $h_i$
denote the width and height of the region $i$, respectively.
After that, for a given lung CT image, we generate a boundary map that separates
the lung and its peripheral tissue, then an image that only contains the lung can be achieved
based on the boundary map.
To generate a boundary map, we slide a 2D kernel with sizes $3\times3$ pixel by pixel on the input lung image.
If a window contains more than one distinct pixel value, the center pixel of this window is marked as a boundary pixel.
At last, we compute the five location tuples for the achieved lung image based on the location tuples $(x_i, y_i, w_i, h_i)$,
where $i\in\{ru, rm, rl, lu, ll\}$. This is because the position and spatial sizes of each region for the human lung are roughly fixed.
We then use the crop operator to crop out five regions based on the location tuples.
Finally, the resize operator is performed such that each region is resized to the original lung CT image's sizes.

We employ either of the above two groups of networks to generate a
region representation for each region.
In this way, the networks are forced to understand and learn inherent
knowledge for each lobe of a lung.
After that, classifiers are used to predict the positions of all the five lobes.
It's essentially a multi-task learning problem. The input is the five regions, each of which covers a lobe.
For a region, the right class is the predefined index for it.
Formally, given an input lung CT image $\bm{X}$, the region-aware loss $\mathcal{L}_{ra}$ for region-scale learning is computed by
\begin{equation}\label{eq:ra_loss}
\mathcal{L}_{ra} = \sum_{i\in\{ru, rm, rl, lu, ll\}}\mbox{CE}(\bm{x}_i, \bm{y}_i),
\end{equation}
where $\mbox{CE}(\bm{x}_i, \bm{y}_i)$ is the cross-entropy
loss for the region $i\in\{ru, rm, rl, lu, ll\}$,
$\bm{x}_i\in\mathbb{R}^5$ is the output of the classifier,
and $\bm{y}_i\in\mathbb{R}^5$ is a one-hot vector indicating
the right class for the region $i$.
Notably, the parameters
are shared across all the five input regions.
The proposed region-scale learning task
has two advantages. Firstly, it learns specific patterns for each lobe, thereby extracting subtle
information at the lobe-level. Secondly, by treating each lobe as a center lobe,
the other four lobes can be viewed as the context for the center lobe.
Thus the center lobe is highlighted and doing well on this task
requires distinguishing representation of each lobe.

\subsection{Final Loss}
As introduced in Section~\ref{sec:imagesl},
an input sample for our self-supervised learning framework contains a CT image
 $\bm{X}$ to form a positive pair,
and a set of different CT images $\{\bm{X}^i|i=1,...,n\}$ to form negative pairs.
Formally, the final loss $\mathcal{L}$ for this input sample
is computed by performing the weighted sum on the contrastive loss of the image-scale learning
and the proposed region-aware loss of the region-scale learning as
\begin{equation}\label{eq:final}
\mathcal{L} = \alpha_1\mathcal{L}_{cons}+\alpha_2(\mathcal{L}_{ra}+\sum^n_{i=1}\mathcal{L}_{ra}^i),
\end{equation}
where $\alpha_1$ and $\alpha_2$ are hyper-parameters, $\mathcal{L}_{cons}$ is defined
in Equation~\ref{eq:contra1}, $\mathcal{L}_{ra}^i$ is the region-aware loss for the image $\bm{X}$,
and $\mathcal{L}_{ra}^i$ is the region-aware loss for an image $\bm{X}^i$ from the image set $\{\bm{X}^i|i=1,...,n\}$.

During back-propagation, the network is forced to
learn representations at both the lung-level and lobe-level,
distinguishing the functionality of each lobe and
capturing sufficient semantic information to achieve better representations
for lung CT images. Note that we use the same network in two branches.
That is to say, the weights are shared for the feature extractors used
in both branches.
Similar to supervised transfer learning, we apply each of the two groups of networks
as the backbone network in the self-supervised learning framework and then transfer the
knowledge to the target task. The transferability of ATTNs and convolution layers
on the self-supervised learning is then
explored.

\subsection{Transferability Estimation} \label{sec:transferability}
We use two approaches to estimate the transferability of learned representations.
We denote a network that is not pretrained on a source task as $N_n$, and the same
network that is pretrained on a source task as $N_p$.
First, we directly finetune the pretrained $N_n$ and $N_p$
on the target task and record the prediction
performance. The metrics include accuracy, F1-score and AUC.
For any metric, the performance of $N_n$ and $N_p$ is denoted
as $P_n$ and $P_p$, respectively.
Then the divergence of each metric between $P_n$ and $P_p$
can be used to estimate the transferability of representations
learned by the employed network on the source task.

In addition to evaluating transferability through running
experiments on the target task, we employ LEEP~\cite{nguyen2020leep} to directly
estimate transferability based on the pretrained model and
statistics of the target dataset.
LEEP can only be applied to supervised transfer learning where
the source data has labels.
Assume labels of the source data are in a label set $\mathcal{Z}$,
input instances of the target data are in  the domain set $\mathcal{X}\in\mathbb{R}^N$,
and labels of the target data are in a label set $\mathcal{Y}$.
Given a pretrained network $N$ and target dataset $\mathcal{D}=\{(x_1, y_1),(x_2, y_2),...,(x_n, y_n)\}$,
where $n$ is the number of data samples in the target set,
$x_i\in\mathcal{X}$ is achieved by flatting an image to a vector.
Formally, the LEEP score $L$ can computed as
\begin{equation}\label{eq:leep}
L = \frac{1}{n}\sum_{i=1}^n\mbox{log}(\sum_{z\in\mathcal{Z}} P(y_i|z)N(x_i)_z),
\end{equation}
where $P(y_i|z)$ is the conditional distribution of the target label $y_i$ given
the source label $z$, and $N(x_i)_z$ is the probability that the output of the network that
takes $x_i$ as the input is the label $z$.
Similarly, the LEEP score based on $N_n$ and $N_p$
with the same target dataset can be denoted as $L_n$ and $L_p$, respectively. The
divergence between $L_n$ and $L_p$ can be used to estimate transferability for
$N_n$ and $N_p$ from the same source task.

\section{Experimental Studies}

\subsection{Dataset}
We use two datasets to pretrain models and perform supervised transfer learning.
First, We use ImageNet that contains millions of natural images to pretrain the models.
Even though nature images follow different distributions with CT images, pretraining on ImageNet
enables models to learn redundant patterns that are shared by natural and medical images.
After that, we use COVIDx~\cite{wang2020covid}, a collection of images from medical domain,
to extract similar patterns that are shared by medical images.
We use COVID19-CT to perform self-supervised transfer learning and CT-based diagnosis of
COVID-19. To our best knowledge, the dataset is the largest public-available CT dataset for COVID-19.

\noindent \textbf{COVIDx} The COVIDx dataset is a public-available labeled dataset containing chest X-ray (CXR)
images for COVID-19 detection. The dataset contains 16898 images in total, among which 573 images are for
COVID-19 cases, 5559 images are for regular pneumonia (non COVID-19) cases and the rest 8066 are normal cases.
The dataset is generated from 5 sources; those are,  COVID-19 Image Data Collection~\cite{cohen2020covid},
COVID-19 Chest X-ray Dataset Initiative~\cite{bworld}, ActualMed COVID-19
Chest X-ray Dataset Initiative~\cite{actualmed}, RSNA Pneumonia Detection Challenge
dataset~\cite{kaggle}
and COVID-19 radiography database~\cite{xxxyyy}.
Generally, CXR imaging is a low-cost, first-look technique compared with
CT scanning. CXR images usually have lower quality than CT images but can
be obtained much faster. Due to their easy-obtaining nature, the existing CXR datasets
are much larger than the CT datasets for COVID-19 diagnosis. However, the CXR imaging and
CT scanning techniques have something in common. The CXR imaging uses a small amount of
radiation to go through and take an image of the chest. CT scanning is essentially
a more detailed type of CXR that makes more comprehensive views of the chest.
In this sense, the achieved CXR and CT images follow similar distributions and may
share common patterns for image representation learning.
It is a vital step to learn and transfer knowledge to the
target dataset from a source dataset that is larger but follows
similar distributions with the target dataset.

\noindent \textbf{COVID19-CT} The COVID19-CT dataset is the largest public-available
CT dataset for COVID-19 diagnosis. It contains 349 images as COVID-19 positive and 397
images as COVID-19 negative. The dataset is originally split into training, validation
and test sets. Specifically, there are 191 positive and 234 negative images in the training set;
60 positive and 58 negative images in the validation set; 98 positive and 105 negative
images in the test set. The images are of different spatial sizes, which vary from
153 to 1853.

\subsection{Experimental Setup}
We employ two backbone networks ResNet-50 and ResNet-101, where purely convolutional layers are used
to extract features from images.
We then investigate two scenarios that adding 1 ATTN and 5 ATTNs
for each backbone network. The ATTN is inserted in the $\mbox{res}_3$ block for the former.
For the latter, 2 ATTNs are inserted in both the $\mbox{res}_3$ and $\mbox{res}_4$,
and another ATTN is inserted in the $\mbox{res}_5$ block. This results in a total of six networks.
Each of the ResNet-50 and ResNet-101 has three variants, including the baseline, the baseline with
1 ATTN and the baseline with 5 ATTNs.
Notably, all ATTNs are added at the end of the corresponding residual
blocks.

We first perform STL-N and pretrain all the six networks on ImageNet ILSVRC
2012 image classification dataset~\cite{imagenet_cvpr09}, which contains 1.2 million natural images for training, 50 thousand for validation
and another 50 thousand for testing. There are 1000 classes in total.
We adopt the same data augmentation pipeline as in~\cite{he2016deep}.
Specifically, each image is scaled to $256\times 256$ and a patch of size $224\times 224$
is randomly cropped as a training sample.
Horizontal flip is randomly performed for each cropped patch with a probability of 0.5.
We employ the dropout~\cite{srivastava2014dropout} with a rate of 0.8 and the weight decay of 1e-4 to avoid over-fitting.
To optimize the models, we employ the stochastic
gradient descent (SGD) optimizer with a momentum of 0.9 to train models for 90 epochs.
The initial learning rate is set to 0.1 and decays by 0.1 every 30 epochs.
We use 8 TITAN Xp GPUs and the batch size is set to 512 for training.

We then perform STL-M to pretrain the models on the COVIDx dataset.
The used data augmentation scheme is composed of several techniques
including translation, rotation, horizontal flip, intensity shift.
All the techniques are randomly performed with a probability of 0.5.
SGD is used with a learning rate
equal to 2e-4. A dropout with a keep rate of 0.5 is adopted to avoid over-fitting.
The number of epochs is set to 30 and the batch size is 64 during the pretraining procedure.

During the SSTL-M, we follow the same data
augmentation scheme as in~\cite{he2020momentum}. Specifically, all images as well as cropped regions are scaled
to $256\time 256$ and a patch with spatial sizes $224\time 224$ is randomly cropped from
each image. Then color jittering, horizontal flip, and grayscale conversion
are performed randomly with a probability of 0.5. The temperature parameter $\tau$ in Equation~\ref{eq:contra1}
is set to 0.07. The weights in Equation~\ref{eq:final} are set to $\alpha_1 = 0.8$, and $\alpha_2 = 0.8$.
The SGD is employed as our optimizer with weight decay equal to 1e-4
and SGD momentum equal to 0.9. The training is performed
for 200 epochs. The initial learning rate is 0.3 and
decays by 0.1 at the 120th and the 160th epochs. The batch size is set to 256 with 8 TITAN Xp GPUs.
When comparing our method with the MoCo and SimCLR, we use the same hyperparameters as used in their papers~\cite{he2020momentum,chen2020simple}
for the MoCo and SimCLR.

After finishing the pretraining procedures, we use each pretrained model
as a starting point and finetune it on the target dataset COVID19-CT for
COVID-19 prediction. We use the same optimizer and hyperparameters as in the SSTL-M.
The data augmentation scheme is also the same during training.
During inference, the center-cropped patch with size $224\time 224$ for each image is used,
and other augmentation techniques are the same.

\begin{table}
\begin{center}
\caption{Top-1 and Top-5 accuracies (\%) and performance improvements of all the six networks on
ImageNet. There are two columns for each metric. The value column provides the original
results, and the improv. column shows the improvements based on the baselines.} \label{tb:imagenet}
\begin{tabular}{lcccccccccccc cc}
\hline
\hhline {--------------} \multicolumn{2}{c}{\multirow{2}*{Model}} & \multicolumn{2}{c}{Top-1} &  \multicolumn{2}{c}{Top-5}  \\
 & &Value & Improv. &Value & Improv. \\
\hhline {--------------} \multirow{3}*{R-50} & Baseline & 77.2  & 0& 93.3 &0 \\
&  1 ATTN & 77.9 & 0.7&  93.9 & 0.6\\
& 5 ATTNs & 78.3 & 1.1& 94.1 & 0.8\\
\hhline {--------------} \multirow{3}*{R-101} & Baseline & 78.3 & 0 & 94.0 & 0 \\
&  1 ATTN & 79.2 & 0.9 & 94.4 & 0.4\\
&  5 ATTNs & 79.5 & 1.2 &94.6 & 0.6\\
\hhline {--------------}
\end{tabular}
\end{center}
\end{table}

\begin{table*}
\begin{center}
\caption{Overall performance for COVID-19 CT image prediction
of all the six models pretrained with
STL-N, STL-M and SSTL-M in terms of accuracy(\%), F1-score(\%) and AUC(\%).
There are two columns for each metric. The value column provides the original results, and the improv. column shows the improvements based on the baselines.
} \label{tb:finalperf}
\begin{tabular}{lcccccccccccc cc}
\hline
\hhline{--------------} \multicolumn{2}{c}{\multirow{2}*{Model}} & &\multicolumn{2}{c}{Accuracy} & &\multicolumn{2}{c}{F1-Score} & &\multicolumn{2}{c}{AUC} \\
 & &&Value & Improv.& &Value & Improv. & &Value & Improv. \\
\hhline {--------------} \multirow{3}*{R-50} & Baseline && 88.2  & 0& &88.7 &0 & &90.3 & 0\\
&  1 ATTN & &93.1 & 4.9 & & 92.9 & 4.2& & 92.9& 2.6\\
&  5 ATTNs & &93.9 & 5.7 & & 94.7 & 6.0& &96.8& 6.5\\
\hhline {--------------} \multirow{3}*{R-101} & Baseline & &89.3 & 0& &88.8 & 0& &90.7 &0\\
&  1 ATTN & &93.4 & 4.1& &93.2 & 4.4& &93.7 & 3.0\\
&  5 ATTNs & &94.2 & 4.9& &95.3 & 6.5& &97.8 & 7.1\\
\hhline {--------------}
\end{tabular}
\end{center}
\end{table*}

\subsection{Power of the Networks with Attention}
We first investigate how powerful the networks with ATTNs are
on ImageNet. We denote ResNet-50 as R-50 and ResNet-101 as R-101 for convenience.
Each of the R-50 and R-101 has three variants including the baseline CNN,
the baseline with 1 ATTN, and the baseline with 5 ATTNs.
We denote them as baseline, 1 ATTN and 5 ATTNs for convenience.
We conduct experiments to predict the top-1 and top-5 accuracies on the validation set
of the ImageNet as the test data has no labels.
In addition, we also compute performance improvements
compared with the baselines. The results are reported in Table~\ref{tb:imagenet}.
We can observe from the table that by adding ATTNs,
the performance improvement is less than or around 1\%
on the validation set of the ImageNet.

We then conduct experiments to examine the overall performance
of all the models on the target dataset.
These models are all pretrained on the three source tasks STL-N, STL-M and SSTL-M in order.
The metrics include accuracy, F1-score and AUC.
For each metric,
we also compute performance improvements compared with the baselines.
The results are reported in Table~\ref{tb:finalperf}.
We can find from the table that adding ATTNs to the networks and then
performing transfer learning can significantly improve prediction performance.
In terms of accuracy, adding 1 ATTN for transfer learning leads to an average
improvement of 4.5\%, and adding 5 ATTNs leads to an average
improvement of 5.3\% compared with the baseline ResNets.
Similar results can be found on the metrics F1-score and AUC.
These results indicate the effectiveness of integrating ATTNs
into our proposed multi-stage transfer learning framework.


\begin{table}
\begin{center}
\caption{Comparison among different self-supervised learning methods in terms of accuracy(\%), F1-score(\%) and AUC(\%).
All the networks are first pretrained on the same supervised source tasks STL-N and STL-M, then pretrained with
different SSTL-M methods on the same CT data.
} \label{tb:ssl}
\begin{tabular}{lcccccccccccc cc}
\hline
\hhline{--------------} \multicolumn{2}{c}{Model} & Accuracy & F1-Score & AUC \\
\hline\hline \multirow{9}*{R-50} & Baseline w MoCo & 87.3  & 88.6 &90.0 \\
& Baseline w SimCLR & 87.7  & 87.9 &90.1 \\
& Baseline w  Ours & 88.2  & 88.7 &90.3 \\
\cline{2-10}&  1 ATTN w MoCo & 92.5 &  92.3 & 92.6\\
&  1 ATTN w SimCLR& 92.6 &  91.7 & 92.1\\
&  1 ATTN w  Ours& 93.1 &  92.9 & 92.9\\
\cline{2-10}&  5 ATTNs w MoCo& 93.6 & 94.3 &96.1\\
&  5 ATTNs w SimCLR& 93.5 & 94.0 &95.9\\
&  5 ATTNs w  Ours& 93.9 & 94.7 &96.8\\
\hline\hline \multirow{9}*{R-101} & Baseline w MoCo& 88.6 &88.3 &90.6\\
& Baseline w SimCLR& 89.1 &88.3 &90.5\\
& Baseline w  Ours& 89.3 &88.8 &90.7\\
\cline{2-10}&  1 ATTN w MoCo& 92.5 &93.0 &93.7\\
&  1 ATTN w SimCLR& 92.6 &92.7 &93.0\\
&  1 ATTN w  Ours& 93.4 &93.2 &93.7\\
\cline{2-10}&  5 ATTNs w MoCo& 93.7 &95.0 &96.6 \\
&  5 ATTNs w SimCLR& 93.1 &94.5 &96.6 \\
&  5 ATTNs w  Ours& 94.2 &95.3 &97.8 \\
\hline\hline
\end{tabular}
\end{center}
\end{table}

\subsection{Comparison of Different Self-supervised Learning Methods}
We investigate the effectiveness of our proposed self-supervised learning method.
For both the R-50 and R-101 groups, we first conduct the same
supervised transfer learning tasks STL-N and STL-M.
After that, we applied the trained models to two state-of-the-art
self-supervised learning frameworks, the MoCo~\cite{he2020momentum} and the SimCLR~\cite{chen2020simple},
and our method, which we denote as X w MoCo, X w SimCLR and X w Ours,
respectively. X denotes either baseline, adding 1 ATTN or adding 5 ATTNs.
The experimental results are reported in Table~\ref{tb:ssl}. We can observe
from the table that models with our method consistently outperform models
with the MoCo or SimCLR on all the three metrics.
Specifically, considering all the three networks together, our method outperforms the MoCo and SimCLR by an average margin of
0.6\% and 0.5\% in terms of accuracy for the R-50 Group, and
an average margin of
0.7\% and 0.6\% in terms of accuracy for the R-101 Group.
Similar results can also be achieved for the F1-score and AUC.
This indicates by using a multi-scale learning framework and
considering distinguishing patterns from local lobes, our
method can successfully extract useful inherent patterns from the CT
data, thereby leading to performance improvement for COVID-19 diagnosis
based on CT images.

\begin{table*}
\begin{center}
\caption{Results for COVID-19 CT image prediction
of all the six networks with and without transfer learning
in terms of accuracy(\%), F1-score(\%) and AUC(\%).
There are two columns for each metric. The value column provides the original results, and the improv. column
shows the performance improvements of networks with transfer learning compared with
networks without transfer learning.} \label{tb:comptl}
\begin{tabular}{lcccccccccccc cc}
\hline
\hhline{--------------} \multicolumn{2}{c}{\multirow{2}*{Model}} &&  \multicolumn{2}{c}{Accuracy} & &\multicolumn{2}{c}{F1-Score} & &\multicolumn{2}{c}{AUC} \\
 & & &Value & Improv.& &Value & Improv. & &Value & Improv. \\
\hline\hline \multirow{6}*{R-50} & Baseline w/o TL & & 83.7 & 0 & &84.1 & 0  && 84.6 & 0\\
& Baseline w TL& & 88.2 & 4.5& &88.7 & 4.6& &90.3 &5.7 \\
\cline{2-11}& 1 ATTN w/o TL & & 84.9 & 0&& 82.4 &0 && 85.7 & 0\\
&  1 ATTN w TL & & 93.1 &8.2 && 92.9 &10.5 && 92.9 &7.2 \\
\cline{2-11}&  5 ATTNs w/o TL & & 84.7 & 0& &83.1 &0 && 86.9 &0 \\
&  5 ATTNs w TL & & 93.9 &9.2 && 94.7 &11.6 && 96.8 &9.9 \\
\hline\hline \multirow{6}*{R-101} & Baseline w/o TL  &&  83.8 & 0&& 83.9 & 0& &86.2 &0 \\
& Baseline w TL & & 89.3 & 5.5& &88.8 & 4.9& &90.7 & 4.5\\
\cline{2-11}&  1 ATTN w/o TL & & 85.4 & 0&& 83.6 &0 && 86.1 & 0\\
&  1 ATTN w TL & & 93.4 & 8.0&& 93.2 & 9.6&& 93.7 & 7.6\\
\cline{2-11}&  5 ATTNs w/o TL & & 86.1 &0 && 84.5 & 0&& 87.6 &0 \\
&  5 ATTNs w TL& & 94.2 & 8.1& &95.3 & 10.8&&97.8 & 10.2\\
\hline\hline
\end{tabular}
\end{center}
\end{table*}

\begin{table*}
\begin{center}
\caption{Results for COVID-19 CT image prediction
of all the six networks without transfer learning, with STL-N and with STL-M
in terms of accuracy(\%), F1-score(\%) and AUC(\%).
There are two columns for each metric. The value column provides the original results, and the improv. column
shows the performance improvements of networks with each of the STL-N and STL-M compared with
networks without transfer learning.} \label{tb:ablat}
\begin{tabular}{lcccccccccccc cc}
\hline
\hhline{--------------} \multicolumn{2}{c}{\multirow{2}*{Model}} & &\multicolumn{2}{c}{Accuracy} & &\multicolumn{2}{c}{F1-Score} & &\multicolumn{2}{c}{AUC} \\
 & &&Value & Improv.& &Value & Improv. & &Value & Improv. \\
\hline\hline \multirow{9}*{R-50} & Baseline w/o TL& &83.7 & 0 && 84.1 & 0 && 84.6 & 0 \\
& Baseline w STL-N  & &85.4 & 1.7 && 85.9 &1.8 && 86.4 &1.8 \\
& Baseline w STL-M & &86.0 & 2.3 & &86.8 & 2.7& &87.7 & 3.1\\

\cline{2-11}&  1 ATTN w/o TL & &84.9 & 0& &82.4 & 0& &85.7 &0 \\
&  1 ATTN w STL-N  & &88.4& 3.5& &88.8 &6.4 & &89.2 & 3.5\\
&  1 ATTN w STL-M & &89.2 & 4.3& &89.8 &7.4 && 92.1 &6.6 \\
\cline{2-11}&  5 ATTNs w/o TL & &84.7 & 0&& 83.1 & 0&& 86.9 & 0\\
&  5 ATTNs w STL-N  & &89.2 &4.5 && 89.6 & 6.5&& 90.2 & 3.3\\
&  5 ATTNs w STL-M &&88.7 &4.0 && 88.8 & 5.7& &91.2 & 4.3\\
\hline\hline \multirow{9}*{R-101} & Baseline w/o TL & &83.8 & 0&& 83.9 & 0&& 86.2 &0 \\
& Baseline + STL-N  & &85.8 & 2.0&& 86.9 & 3.0&& 88.1 &1.9 \\
& Baseline + STL-M & &86.1 & 2.3&& 86.8 & 2.9&& 88.5 & 2.3\\

\cline{2-11}&  1 ATTN w/o TL & &85.4 & 0&& 83.6 &0 && 86.1 & 0\\
&  1 ATTN w STL-N  & &88.3 & 2.9&& 89.2 &5.6 && 90.1 &4.0\\
&  1 ATTN w STL-M & &89.5 & 3.1& &89.7 &6.1 && 91.8 & 5.7\\
\cline{2-11}&  5 ATTNs w/o TL& &86.1 & 0& &84.5 & 0& &87.6 & 0\\
&  5 ATTNs w STL-N  & &89.4 & 3.3&& 89.6 & 5.1&& 92.2 &4.6 \\
& 5 ATTNs w STL-M & &90.0 & 3.9& &90.2 & 5.7& &92.4 & 4.8\\
\hline\hline
\end{tabular}
\end{center}
\end{table*}

\subsection{Attention Benefits Transfer Learning}
We design experiments to explore whether ATTNs bring benefits
for transfer learning. For both the R-50 and R-101 groups we conduct
two sets of experiments.
First, we directly optimize the networks on the target dataset COVID19-CT without
transfer learning, namely X w/o TL, where X denotes either baseline, adding 1 ATTN or adding 5 ATTNs.
Second, we apply all the networks to our multi-stage transfer learning framework
that we
perform all the three stages of pretraining for all the networks and then
finetune the pretrained models on the target dataset. We name such models as X w TL,
where X has the same meaning as above.
The performance is evaluated on the test set of the COVID19-CT dataset.
We then compute the improvements by using transfer learning
for the two groups of networks and the results are reported in Table~\ref{tb:comptl}.
We can observe from the table that models with the proposed transfer learning framework consistently improve
the prediction performance on the target dataset. This indicates
that by using our multi-stage transfer learning framework,
it successfully extracts
important patterns between the source images and the target images.

More importantly, the table shows that networks with ATTNs achieve
much larger performance improvements through transfer learning than
the baseline ResNets.
In terms of accuracy on ResNet-50, the improvement for the baseline is 4.5\%,
while and improvements for adding 1 ATTN and adding 5 ATTNs are 8.2\% and 9.2\%, respectively.
The benefits induced by attention are 2.7\% and 3.7\%.
Similar results can be observed for the ResNet-101 and the
benefits induced by attention are 2.5\% and 2.6\%.
Consistent conclusions can be obtained on the metrics F1-score and AUC that
attention helps transfer learning.
For the two settings with ATTNs, the benefits brought by attention are 5.9\% and 7.0\% in terms of
F1-score for ResNet-50, and 4.7\% and 5.9\% for ResNet-101.
In terms of AUC, attention helps with margins of 2.5\% and 4.2\% for ResNet-50,
and margins of 3.1\% and 5.7\% for ResNet-101.
These results indicate that compared with convolution, attention helps transfer more useful knowledge from
the source tasks to the target task in transfer learning. By adding ATTNs, the network is capable of
learning important common pattens of images in the pretraining stage, thereby
leading to significant performance improvement for the target task.

\begin{table}[!bh]
\begin{center}
\caption{LEEP scores for all the six networks without transfer learning, with STL-N, with both the STL-N and STL-M.} \label{tb:leep}
\begin{tabular}{lc ccccccccccc cc}
\hline
\hhline{--------------}  Model& R-50 & R-101 \\
\hhline{--------------}  Baseline w/o TL & -0.918 & -0.913 \\
Baseline w STL-N  & -0.908 & 0.-909 \\
Baseline w STL-N w STL-M & -0.894 & -0.899  \\
\hhline{--------------}
  1 ATTN w/o TL & -0.917 & -0.904 \\
 1 ATTN w STL-N  & -0.902 & -0.887 \\
 1 ATTN w STL-N w STL-M & -0.882 & -0.872 \\
\hhline{--------------}
  5 ATTNs w/o TL & -0.915 & -0.907  \\
  5 ATTNs w STL-N  & -0.897 & -0.880 \\
  5 ATTNs w STL-N w STL-M & -0.875 & -0.865 \\
\hline
\end{tabular}
\end{center}
\end{table}

\begin{figure*}[!th]
\centering
\includegraphics[width=0.9\textwidth]{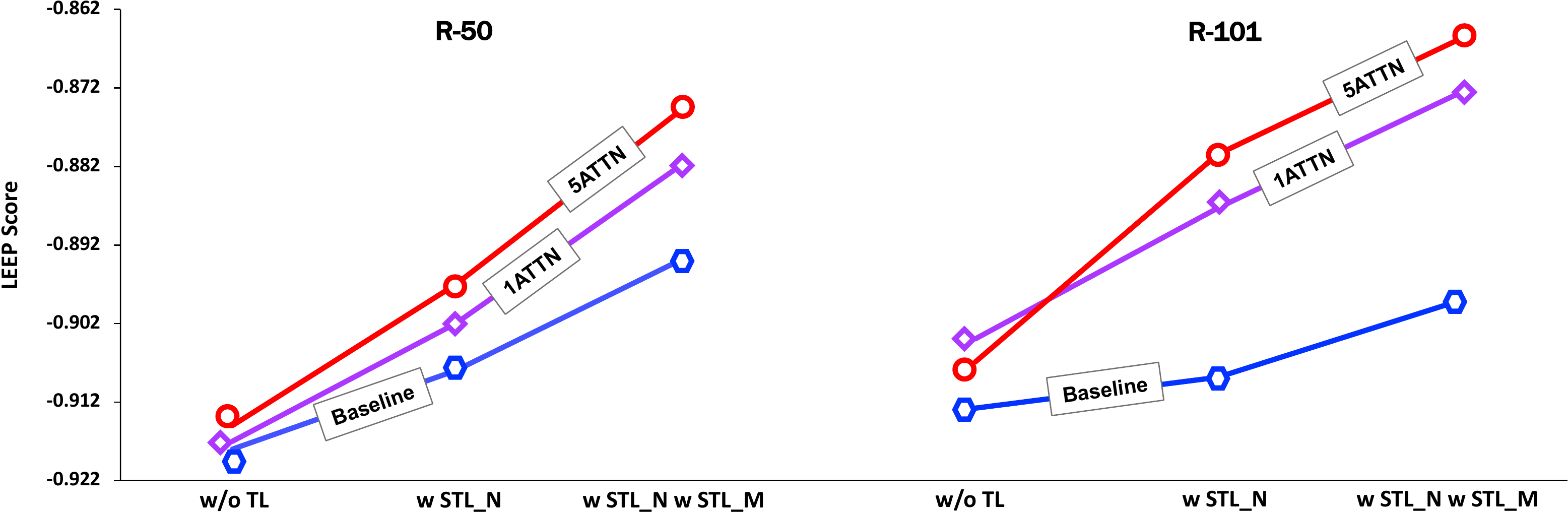}
\caption{LEEP scores for both the R-50 and R-101 groups without transfer learning, with STL-N, with both the STL-N and STL-M.} \label{fig:leep}
\end{figure*}

\begin{figure*}[!th]
\centering
\includegraphics[width=0.9\textwidth]{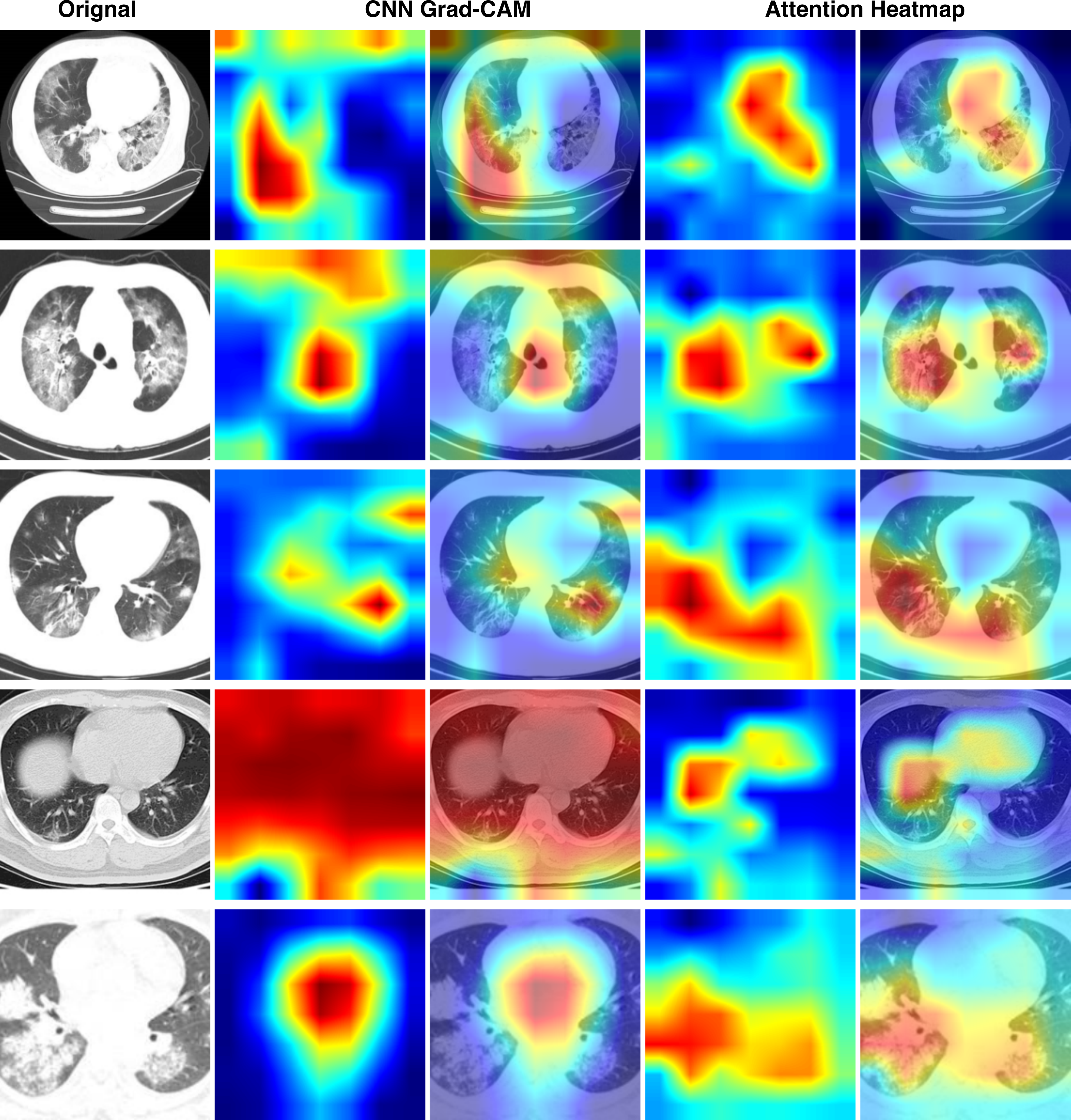}
\caption{Comparison of visualization results between a convolutional
layer and a ATTN. The lowest pixel values are shown in blue while
the highest pixel values are shown in red.
Column 1 are the original CT images with
COVID-19 positive. Columns 2 and 3 are
visualization results for the last convolutional layer in ResNet-50,
where column 2 are the generated heatmaps and column 3 are the original images
integrated with heatmaps.
Columns 4 and 5 are
visualization results for the last ATTN in ResNet-50 with 5 ATTNs,
where column 4 are the generated attention heatmaps and column 5 are the original images
integrated with attention heatmaps.
} \label{fig:vis}
\end{figure*}

\subsection{Ablation Study} \label{sec:ablation}
In this section, we conduct ablation study to examine the benefits of attention brought
to the two supervised transfer learning tasks STL-N and STL-M.
Instead of performing two tasks sequentially,
we conduct pretraining on natural images and medical images
separately to explore how attention benefits transfer learning from different domains.
Specifically, we name the models as X w/o TL, X w STL-N and X w STL-M, respectively. X denotes either baseline, adding 1 ATTN or adding 5 ATTNs.
We compute the improvements for each stage of transfer learning
for both the R-10 and R-101 groups and the results are reported in Table~\ref{tb:ablat}.
We can observe from the table that attention consistently benefits both
the STL-N and STL-M.
Specifically, on the ResNet-50, adding 1 ATTN benefits
the STL-N by a margin of 1.8\% in terms of accuracy, 4.6\% in terms of F1-score,
and 1.7\% in terms of AUC. Adding 1 ATTNs benefits
the STL-M by a margin of 2.0\% in terms of accuracy, 4.7\% in terms of F1-score,
and 3.5\% in terms of AUC.
On the ResNet-50, adding 5 ATTNs benefits
the STL-N by a margin of 2.8\% in terms of accuracy, 4.7\% in terms of F1-score,
and 1.5\% in terms of AUC. Adding 5 ATTNs benefits
the STL-M by a margin of 1.7\% in terms of accuracy, 3.0\% in terms of F1-score,
and 1.3\% in terms of AUC.
Similar results can be computed on ResNet-101 that
attention consistently benefits both the STL-N and STL-M on all three metrics.
These results indicate that attention helps transfer much more useful knowledge
than convolution in supervised transfer learning, no matter the source data follows
totally different or similar distributions with the target data.

\subsection{Comparison on LEEP Scores}
In this section, we compute LEEP scores to examine the functionality
of attention in transfer learning. Instead of optimizing parameters and making
predictions on the target dataset, we directly compute LEEP scores
based on the source models and statistics of the target data.
As LEEP scores can only be computed for supervised transfer learning, we conduct experiments
on STL-N and STL-M to achieve LEEP scores. we add STL-N and STL-M in order, which results in models
X w/o TL, X w STL-N and X w STL-N w STL-M, respectively. X denotes either baseline, adding 1 ATTN or adding 5 ATTNs.
The results are reported in Table~\ref{tb:leep} and shown in Figure~\ref{fig:leep}.
The curve for each network setting in Figure~\ref{fig:leep} is composed of two line segments,
each of which indicates improvement by adding a stage of pretraining procedure.
The slope of each line segment can reflect the improvement in the LEEP score by
adding the corresponding stage of transfer learning.
We can observe from the figure that both adding 1 ATTN and adding 5 ATTNs achieve
larger improvement than the baseline ResNets for all the two stages.
This again demonstrates attention helps transfer learning
regardless of
divergences between the distributions of the
source datasets and the target dataset.

\subsection{Qualitative Results}
In addition to the quantitative results in the above sections, we
provide qualitative results to show the capability of our proposed multi-stage attentive transfer learning framework when detecting
important regions in CT image for COVID-19 diagnosis.
We use ResNet-50 as the baseline and then adding 5 ATTNs to the baseline.
Both the networks are pretrained on the three source tasks and finetuned on the target task.
We use CNN Grad-CAM~\cite{selvaraju2017grad} to visualize convolution maps
for the last convolutional layer right before the global average pooling.
For the network with 5 ATTNs, we visualize the last ATTN inserted in the $\mbox{res}_5$.
We simply perform average on attention score maps of all the pixels
and achieve the final attention map.
As shown in Figure~\ref{fig:vis}, after performing transfer learning, a ATTN layer
can successfully detect regions that are severely infected by the virus. However,
convolutional layers fail in some cases where the uninfected regions are highlighted.
This again demonstrates the effectiveness of using attention in our proposed multi-stage transfer learning framework.

\section{Conclusion}
We propose a unified transfer learning framework and examine how attention facilitates transfer learning for improving COVID-19 diagnosis.
We first design a multi-stage transfer learning framework, which consists of
supervised transfer learning from natural images (STL-N), supervised transfer learning from medical images (STL-M)
and self-supervised transfer learning from medical images (SSTL-M).
This framework allows transferring knowledge from data of different domains,
such as large-scale labeled natural images, large-scale labeled medical images
and the same CT images.
As existing self-supervised learning methods usually generate
poor results on tasks related to medical images,
we propose a novel self-supervised learning method based
on the understanding of substructures of the human lung. The method is integrated as
the last stage in our transfer learning framework
and self-supervised transfer learning is performed
to reuse the complex patterns learned
from the same CT images.
Experimental results show that our method outperforms several SOTA baseline methods.
For the networks used in our transfer learning framework,
we integrate self-attention layers into ResNets and
apply them to the proposed transfer learning framework.
Experimental results
demonstrate attention has higher transferability than convolution. To our best knowledge,
this is the first work to compare transferability of attention and convolution when the source tasks and data are the same in transfer learning.

\ifCLASSOPTIONcompsoc
  \section*{Acknowledgments}
\else
  \section*{Acknowledgment}
\fi

This work was supported by National Science Foundation grants
DBI-1922969 and IIS-1908220.


\begin{thebibliography}{10}
\providecommand{\url}[1]{#1}
\csname url@samestyle\endcsname
\providecommand{\newblock}{\relax}
\providecommand{\bibinfo}[2]{#2}
\providecommand{\BIBentrySTDinterwordspacing}{\spaceskip=0pt\relax}
\providecommand{\BIBentryALTinterwordstretchfactor}{4}
\providecommand{\BIBentryALTinterwordspacing}{\spaceskip=\fontdimen2\font plus
\BIBentryALTinterwordstretchfactor\fontdimen3\font minus
  \fontdimen4\font\relax}
\providecommand{\BIBforeignlanguage}[2]{{%
\expandafter\ifx\csname l@#1\endcsname\relax
\typeout{** WARNING: IEEEtran.bst: No hyphenation pattern has been}%
\typeout{** loaded for the language `#1'. Using the pattern for}%
\typeout{** the default language instead.}%
\else
\language=\csname l@#1\endcsname
\fi
#2}}
\providecommand{\BIBdecl}{\relax}
\BIBdecl

\bibitem{tang2020laboratory}
Y.-W. Tang, J.~E. Schmitz, D.~H. Persing, and C.~W. Stratton, ``Laboratory
  diagnosis of covid-19: current issues and challenges,'' \emph{Journal of
  clinical microbiology}, vol.~58, no.~6, 2020.

\bibitem{udugama2020diagnosing}
B.~Udugama, P.~Kadhiresan, H.~N. Kozlowski, A.~Malekjahani, M.~Osborne, V.~Y.
  Li, H.~Chen, S.~Mubareka, J.~B. Gubbay, and W.~C. Chan, ``Diagnosing
  covid-19: the disease and tools for detection,'' \emph{ACS nano}, vol.~14,
  no.~4, pp. 3822--3835, 2020.

\bibitem{xie2020chest}
X.~Xie, Z.~Zhong, W.~Zhao, C.~Zheng, F.~Wang, and J.~Liu, ``Chest ct for
  typical 2019-ncov pneumonia: relationship to negative rt-pcr testing.''
  \emph{Radiology}, pp. 200\,343--200\,343, 2020.

\bibitem{zhou2020clinical}
F.~Zhou, T.~Yu, R.~Du, G.~Fan, Y.~Liu, Z.~Liu, J.~Xiang, Y.~Wang, B.~Song,
  X.~Gu \emph{et~al.}, ``Clinical course and risk factors for mortality of
  adult inpatients with covid-19 in wuhan, china: a retrospective cohort
  study,'' \emph{The lancet}, 2020.

\bibitem{bernheim2020chest}
A.~Bernheim, X.~Mei, M.~Huang, Y.~Yang, Z.~A. Fayad, N.~Zhang, K.~Diao, B.~Lin,
  X.~Zhu, K.~Li \emph{et~al.}, ``Chest ct findings in coronavirus disease-19
  (covid-19): relationship to duration of infection,'' \emph{Radiology}, p.
  200463, 2020.

\bibitem{li2020coronavirus}
Y.~Li and L.~Xia, ``Coronavirus disease 2019 (covid-19): role of chest ct in
  diagnosis and management,'' \emph{American Journal of Roentgenology}, vol.
  214, no.~6, pp. 1280--1286, 2020.

\bibitem{wang2020covid}
L.~Wang and A.~Wong, ``Covid-net: A tailored deep convolutional neural network
  design for detection of covid-19 cases from chest x-ray images,'' \emph{arXiv
  preprint arXiv:2003.09871}, 2020.

\bibitem{gozes2020rapid}
O.~Gozes, M.~Frid-Adar, H.~Greenspan, P.~D. Browning, H.~Zhang, W.~Ji,
  A.~Bernheim, and E.~Siegel, ``Rapid ai development cycle for the coronavirus
  (covid-19) pandemic: Initial results for automated detection \& patient
  monitoring using deep learning ct image analysis,'' \emph{arXiv preprint
  arXiv:2003.05037}, 2020.

\bibitem{alimadadi2020artificial}
A.~Alimadadi, S.~Aryal, I.~Manandhar, P.~B. Munroe, B.~Joe, and X.~Cheng,
  ``Artificial intelligence and machine learning to fight covid-19,'' 2020.

\bibitem{wynants2020prediction}
L.~Wynants, B.~Van~Calster, M.~M. Bonten, G.~S. Collins, T.~P. Debray,
  M.~De~Vos, M.~C. Haller, G.~Heinze, K.~G. Moons, R.~D. Riley \emph{et~al.},
  ``Prediction models for diagnosis and prognosis of covid-19 infection:
  systematic review and critical appraisal,'' \emph{bmj}, vol. 369, 2020.

\bibitem{wang2020deep}
S.~Wang, B.~Kang, J.~Ma, X.~Zeng, M.~Xiao, J.~Guo, M.~Cai, J.~Yang, Y.~Li,
  X.~Meng \emph{et~al.}, ``A deep learning algorithm using ct images to screen
  for corona virus disease (covid-19),'' \emph{MedRxiv}, 2020.

\bibitem{he2020sample}
X.~He, X.~Yang, S.~Zhang, J.~Zhao, Y.~Zhang, E.~Xing, and P.~Xie,
  ``Sample-efficient deep learning for covid-19 diagnosis based on ct scans,''
  \emph{medRxiv}, 2020.

\bibitem{pan2009survey}
S.~J. Pan and Q.~Yang, ``A survey on transfer learning,'' \emph{IEEE
  Transactions on knowledge and data engineering}, vol.~22, no.~10, pp.
  1345--1359, 2009.

\bibitem{weiss2016survey}
K.~Weiss, T.~M. Khoshgoftaar, and D.~Wang, ``A survey of transfer learning,''
  \emph{Journal of Big data}, vol.~3, no.~1, p.~9, 2016.

\bibitem{yosinski2014transferable}
J.~Yosinski, J.~Clune, Y.~Bengio, and H.~Lipson, ``How transferable are
  features in deep neural networks?'' in \emph{Advances in neural information
  processing systems}, 2014, pp. 3320--3328.

\bibitem{doersch2015unsupervised}
C.~Doersch, A.~Gupta, and A.~A. Efros, ``Unsupervised visual representation
  learning by context prediction,'' in \emph{Proceedings of the IEEE
  international conference on computer vision}, 2015, pp. 1422--1430.

\bibitem{gidaris2018unsupervised}
S.~Gidaris, P.~Singh, and N.~Komodakis, ``Unsupervised representation learning
  by predicting image rotations,'' \emph{arXiv preprint arXiv:1803.07728},
  2018.

\bibitem{lee2019rethinking}
H.~Lee, S.~J. Hwang, and J.~Shin, ``Rethinking data augmentation:
  Self-supervision and self-distillation,'' \emph{arXiv preprint
  arXiv:1910.05872}, 2019.

\bibitem{kolesnikov2019revisiting}
A.~Kolesnikov, X.~Zhai, and L.~Beyer, ``Revisiting self-supervised visual
  representation learning,'' in \emph{Proceedings of the IEEE conference on
  Computer Vision and Pattern Recognition}, 2019, pp. 1920--1929.

\bibitem{zhai2019s4l}
X.~Zhai, A.~Oliver, A.~Kolesnikov, and L.~Beyer, ``S4l: Self-supervised
  semi-supervised learning,'' in \emph{Proceedings of the IEEE international
  conference on computer vision}, 2019, pp. 1476--1485.

\bibitem{devlin2018bert}
J.~Devlin, M.-W. Chang, K.~Lee, and K.~Toutanova, ``Bert: Pre-training of deep
  bidirectional transformers for language understanding,'' \emph{arXiv preprint
  arXiv:1810.04805}, 2018.

\bibitem{ren2015faster}
S.~Ren, K.~He, R.~Girshick, and J.~Sun, ``Faster r-cnn: Towards real-time
  object detection with region proposal networks,'' in \emph{Advances in neural
  information processing systems}, 2015, pp. 91--99.

\bibitem{raghu2019transfusion}
M.~Raghu, C.~Zhang, J.~Kleinberg, and S.~Bengio, ``Transfusion: Understanding
  transfer learning for medical imaging,'' in \emph{Advances in neural
  information processing systems}, 2019, pp. 3347--3357.

\bibitem{esteva2017dermatologist}
A.~Esteva, B.~Kuprel, R.~A. Novoa, J.~Ko, S.~M. Swetter, H.~M. Blau, and
  S.~Thrun, ``Dermatologist-level classification of skin cancer with deep
  neural networks,'' \emph{nature}, vol. 542, no. 7639, pp. 115--118, 2017.

\bibitem{dosovitskiy2015discriminative}
A.~Dosovitskiy, P.~Fischer, J.~T. Springenberg, M.~Riedmiller, and T.~Brox,
  ``Discriminative unsupervised feature learning with exemplar convolutional
  neural networks,'' \emph{IEEE transactions on pattern analysis and machine
  intelligence}, vol.~38, no.~9, pp. 1734--1747, 2015.

\bibitem{he2020momentum}
K.~He, H.~Fan, Y.~Wu, S.~Xie, and R.~Girshick, ``Momentum contrast for
  unsupervised visual representation learning,'' in \emph{Proceedings of the
  IEEE/CVF Conference on Computer Vision and Pattern Recognition}, 2020, pp.
  9729--9738.

\bibitem{chen2020improved}
X.~Chen, H.~Fan, R.~Girshick, and K.~He, ``Improved baselines with momentum
  contrastive learning,'' \emph{arXiv preprint arXiv:2003.04297}, 2020.

\bibitem{chen2020simple}
T.~Chen, S.~Kornblith, M.~Norouzi, and G.~Hinton, ``A simple framework for
  contrastive learning of visual representations,'' \emph{arXiv preprint
  arXiv:2002.05709}, 2020.

\bibitem{wang2018non}
X.~Wang, R.~Girshick, A.~Gupta, and K.~He, ``Non-local neural networks,'' in
  \emph{Proceedings of the IEEE conference on computer vision and pattern
  recognition}, 2018, pp. 7794--7803.

\bibitem{vaswani2017attention}
A.~Vaswani, N.~Shazeer, N.~Parmar, J.~Uszkoreit, L.~Jones, A.~N. Gomez,
  {\L}.~Kaiser, and I.~Polosukhin, ``Attention is all you need,'' in
  \emph{Advances in Neural Information Processing Systems}, 2017, pp.
  6000--6010.

\bibitem{kolda2009tensor}
T.~G. Kolda and B.~W. Bader, ``Tensor decompositions and applications,''
  \emph{SIAM review}, vol.~51, no.~3, pp. 455--500, 2009.

\bibitem{he2016deep}
K.~He, X.~Zhang, S.~Ren, and J.~Sun, ``Deep residual learning for image
  recognition,'' in \emph{Proceedings of the IEEE conference on computer vision
  and pattern recognition}, 2016, pp. 770--778.

\bibitem{chen2019self}
L.~Chen, P.~Bentley, K.~Mori, K.~Misawa, M.~Fujiwara, and D.~Rueckert,
  ``Self-supervised learning for medical image analysis using image context
  restoration,'' \emph{Medical image analysis}, vol.~58, p. 101539, 2019.

\bibitem{drake2009gray}
R.~Drake, A.~W. Vogl, and A.~W. Mitchell, \emph{Gray's Anatomy for Students
  E-Book}, 3rd~ed.\hskip 1em plus 0.5em minus 0.4em\relax Elsevier Health
  Sciences, 2009, pp. 167--174.

\bibitem{nguyen2020leep}
C.~V. Nguyen, T.~Hassner, C.~Archambeau, and M.~Seeger, ``Leep: A new measure
  to evaluate transferability of learned representations,'' \emph{arXiv
  preprint arXiv:2002.12462}, 2020.

\bibitem{cohen2020covid}
J.~P. Cohen, P.~Morrison, and L.~Dao, ``Covid-19 image data collection,''
  \emph{arXiv preprint arXiv:2003.11597}, 2020.

\bibitem{bworld}
I.~Lütkebohle, ``Bworld robot control software,''
  \emph{http://aiweb.techfak.uni-bielefeld.de/content/bworld-robot-control-software},
  2008.

\bibitem{actualmed}
A.~Chung, ``Actualmed {COVID-19} chest x-ray data initiative,''
  \emph{https://github.com/agchung/Actualmed-COVID-chestxray-dataset}, 2020.

\bibitem{kaggle}
R.~S. of~North~America, ``Rsna pneumonia detection challenge,''
  \emph{https://www.kaggle.com/tawsifurrahman/covid19-radiography-database},
  2019.

\bibitem{xxxyyy}
R.~of~North~America, ``{COVID-19} radiography database,''
  \emph{https://www.kaggle.com/c/rsna-pneumonia-detection-challenge/data},
  2019.

\bibitem{imagenet_cvpr09}
J.~Deng, W.~Dong, R.~Socher, L.-J. Li, K.~Li, and L.~Fei-Fei, ``{ImageNet: A
  Large-Scale Hierarchical Image Database},'' in \emph{CVPR09}, 2009.

\bibitem{srivastava2014dropout}
N.~Srivastava, G.~Hinton, A.~Krizhevsky, I.~Sutskever, and R.~Salakhutdinov,
  ``Dropout: a simple way to prevent neural networks from overfitting,''
  \emph{The Journal of Machine Learning Research}, vol.~15, no.~1, pp.
  1929--1958, 2014.

\bibitem{selvaraju2017grad}
R.~R. Selvaraju, M.~Cogswell, A.~Das, R.~Vedantam, D.~Parikh, and D.~Batra,
  ``Grad-cam: Visual explanations from deep networks via gradient-based
  localization,'' in \emph{Proceedings of the IEEE international conference on
  computer vision}, 2017, pp. 618--626.

\end{thebibliography}
\end{document}